\documentclass[twocolumn,floats]{article}

\usepackage{epsf}
\usepackage{times}

\newcommand{\figurewidth}{\linewidth}

\begin{document}

% \draft

\title{
Fluid Particle Accelerations in Fully
Developed Turbulence }

\author{A. La Porta,
Greg A. Voth, Alice M. Crawford,
Jim Alexander, and Eberhard Bodenschatz\\
\textit{Laboratory of Atomic and Solid State Physics,}
\textit{Laboratory of Nuclear Studies}\\
\textit{Cornell University, Ithaca, NY 14853-2501}
}

\date{November 5, 2000}

\maketitle
\thispagestyle{empty}

\pagebreak

\textbf{%
The motion of fluid particles as they are pushed along erratic
trajectories by fluctuating pressure gradients
is fundamental to transport and mixing in turbulence. 
It is essential in cloud formation and atmospheric
transport\cite{vaillancourt:2000,weil:1992}, 
processes in stirred
chemical reactors and combustion systems\cite{pope:1994},
and in the industrial production of nanoparticles\cite{pratsinis:1996}.
The perspective of particle trajectories has been used successfully
to describe mixing and transport in turbulence\cite{pope:1994,shraiman:2000}, 
but issues of fundamental importance remain unresolved.
One such issue is the Heisenberg-Yaglom 
prediction of fluid particle 
accelerations\cite{heisenberg:1948,yaglom:1949}, based on
the 1941 scaling 
theory of Kolmogorov\cite{Kolmogorov:1941:LST,Kolmogorov:1941:DEL} (K41).
Here we report acceleration measurements 
using a detector adapted from high-energy physics to track particles in
a laboratory water flow at Reynolds numbers up to 63,000.
We find that  universal K41 scaling of the acceleration variance is attained
at high Reynolds numbers.
Our data show strong intermittency---particles are observed with
accelerations of up to 1,500 times the acceleration of gravity
(40 times the root mean square value).  
Finally, we find that accelerations manifest the anisotropy
of the large scale flow at all Reynolds numbers studied.}

In principle, fluid particle trajectories are easily
measured by seeding a turbulent flow with minute tracer 
particles and following their motions with an imaging system. 
In practice this can be a very challenging task
since we must fully resolve particle motions
which take place on times scales of the order
of the Kolmogorov time, $\tau_\eta = (\nu/\epsilon)^{1/2}$ where
$\nu$ is the kinematic viscosity and $\epsilon$ is the turbulent
energy dissipation.  
This is exemplified in Fig.~\ref{fig:traj}, which shows a measured
three-dimensional, time resolved trajectory of a tracer particle
undergoing violent accelerations in our turbulent water flow,
for which $\tau_\eta = 0.3\ \mathrm{ms}$. 
The particle enters the detection volume on the upper right,
is pushed to the left by a burst of acceleration
and comes nearly to a stop before being rapidly accelerated (1200 times
the acceleration of gravity) upward in a cork-screw motion.
This trajectory illustrates the difficulty in
following tracer particles---a particle's acceleration can go from zero to 30
times its rms value and back to zero in 
fractions of a millisecond
and within distances of hundreds of micrometers.

\begin{figure}[tb]
\epsfxsize=\linewidth
\begin{center}
\mbox{\epsffile{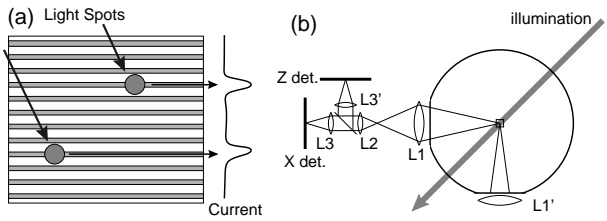}} 
\end{center}
\caption{%
\textsc{Apparatus} (a) Schematic
representation of the CLEO~III strip detector\protect\cite{skubic:1998}, 
in which grey bars
indicate sense strips which collect charge carriers freed by optical radiation.
The 511 strips allow measurement of the one dimensional projection
of the light striking the detector. 
The detector may be read out 70,000 times per second.
(b) A combination of lenses (L1, L2, L3, L3') is used to 
image the active volume onto a pair of strip detectors
which are oriented to measure the $x$ and $y$ coordinates.  
Another detector assembly may be placed on the opposite port (L1') to measure
$y$ and $z$.  
The flow is illuminated by a
6~W argon ion laser beam oriented at $45^\circ$ with respect to the
two viewports. 
The optics image $(46 \pm 7)$~$\mu$m diameter
transparent polystyrene spheres which have a density of 1.06 ${\rm
g/cm}^3$. 
Particle positions are measured with accuracy 0.1 strips, corresponding to
0.7~$\mu$m in the flow. } 
\label{fig:detector}
\end{figure}

Conventional detector technologies are effective for low Reynolds
number flows\cite{virant:1997,ott:2000}, 
but do not provide adequate temporal resolution
at high Reynolds numbers.
However, the requirements are met by 
the use of silicon strip detectors as optical imaging elements
in a particle tracking system.
The strip detectors employed in our experiment (See Fig.~\ref{fig:detector}a) 
were developed to measure
particle tracks in the vertex detector of the CLEO~III experiment
operating at the Cornell Electron Positron Collider\cite{skubic:1998}. 
When applied to particle tracking in turbulence 
(See Fig.~\ref{fig:detector}b) 
each detector measures a one-dimensional projection of the 
image of the tracer particles. 
Using a data acquisition system designed for the turbulence experiment, 
several detectors can be simultaneously read out
at up to 70,000 frames per second.

The acceleration of a fluid particle, $\mathbf{a^+}$, 
in a turbulent flow is given by the
Navier-Stokes equations,
\begin{equation}
\mathbf{a^+} = -\frac{\vec{\nabla} p}{\rho} + \nu \nabla^2 \mathbf{u}
\end{equation}
where $p$ is the pressure, $\rho$ is the fluid density, 
and $\mathbf{u}$ is the velocity field.  
In fully developed turbulence
the viscous damping term is small compared to the pressure
gradient term\cite{batchelor:1951,vedula:1998} and therefore the 
acceleration is closely related to the
pressure gradient.

\begin{figure}[tb]
\epsfxsize=\figurewidth
\begin{center}
\mbox{\epsffile{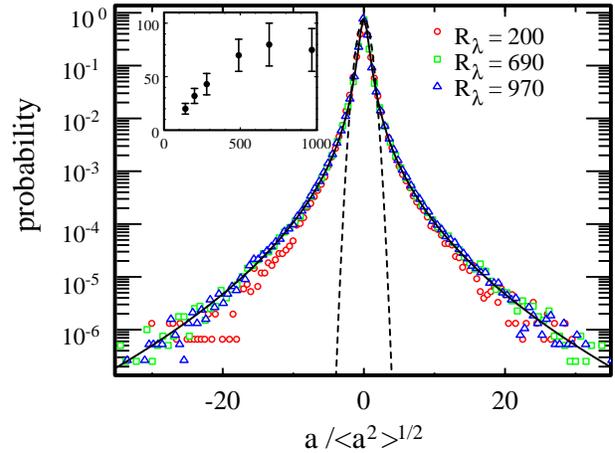}}
\end{center}
\caption{\textsc{Acceleration Distribution.} Probability density functions
of the transverse acceleration normalized by its standard deviation
at different Reynolds numbers.
The acceleration is measured from parabolic fits over
$0.75$~$\tau_\eta$ segments of each trajectory. The solid line is a
parameterization of the highest Reynolds number data using the
function $P(a) = C \exp \left( - a^2 / \left(\left(1 + \left|a \beta
/ \sigma \right|^\gamma \right) \sigma^2 \right)  \right)$, with
$\beta=0.539$, $\gamma = 1.588$, $\sigma = 0.508$ and the dashed line
is a Gaussian distribution with the same variance.
The inset shows the flatness 
of the acceleration distribution, 
($\langle a^4\rangle / \langle a^2 \rangle^2$,
evaluated using
$0.5$~$\tau_\eta$ parabolic fits)
as a function of $R_\lambda$. }
\label{fig:acc_dist}
\end{figure}

Our measurement of the distribution of accelerations is shown in
Figure~\ref{fig:acc_dist}, where the probability density function
of a normalized acceleration component is plotted at three Reynolds
numbers.
All of the distributions have a stretched exponential shape, in which
the tails extend much further than they would for a Gaussian distribution
with the same variance.
This indicates that accelerations many times the rms value
are not as rare as one might
expect, \textit{i.e.}, the acceleration is extremely intermittent.
The acceleration flatness, shown in the inset to Fig.~\ref{fig:acc_dist},
characterizes the intermittency of the acceleration, and would be
3 for a Gaussian distribution.
These flatness values are consistent 
with direct numerical simulation (DNS) at low Reynolds 
number\cite{vedula:1998} and exceed 60
at the highest Reynolds numbers.

The prediction by Heisenberg and Yaglom for the 
variance of an acceleration component based on K41 theory is
\begin{equation}
\label{eq:avar} \langle a_i a_j \rangle = a_0 \epsilon^{3/2}
\nu^{-1/2} \delta_{ij},
\label{eq:scale}
\end{equation}
where $a_0$ is a universal constant which is approximately 1 
in a model assuming Gaussian 
fluctuations\cite{heisenberg:1948,yaglom:1949,obukhov:1951,batchelor:1951}. 
However, DNS has found that $a_0$ depends on $\epsilon$.
Conventionally this is expressed in terms of the
Taylor microscale Reynolds number, $R_\lambda$,
which is related to the conventional Reynolds number by 
$R_\lambda = (15 \mathrm{Re})^{1/2}$
and is proportional to $\epsilon^{1/6}$. 
Using this notation, DNS results indicate
$a_0 \sim R_\lambda^{1/2}$ for
$R_\lambda < 250$\cite{vedula:1998}, with a tendency to level off
as $R_\lambda$ approaches 470\cite{gotoh:2000}.

\begin{figure}[tb]
\epsfxsize=\figurewidth
\begin{center}
\mbox{\epsffile{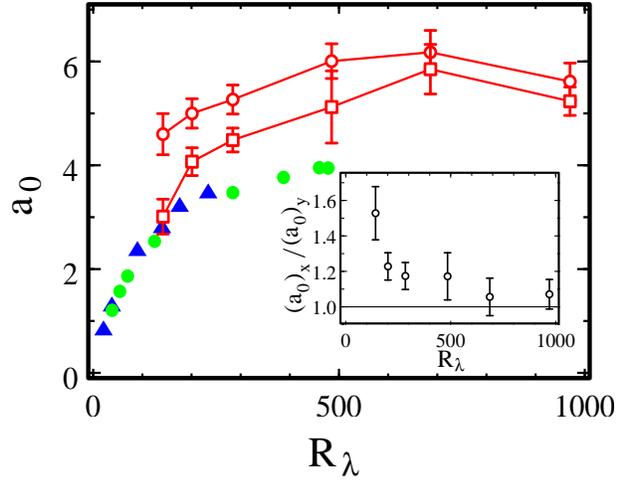}}
\end{center}
\caption{\textsc{$a_0$ as a function of $R_\lambda$.}  Open red circles
indicate a transverse component and open red squares the axial
component of the acceleration variance. 
DNS data is represented by blue triangles\protect\cite{vedula:1998} 
and green circles\protect\cite{gotoh:2000}.
The error bars represent random and systematic errors in the
measurement of the acceleration variance.
There is an additional
uncertainty of 15\% in the overall scaling of the vertical axis for
the experimental data due to the uncertainty in the measured value of
the energy dissipation.  
The degree to which the 45~$\mu$m diameter tracer particles follow
the flow was investigated by measuring 
the acceleration variance as a function of particle size and
density.
The results, to be published elsewhere,
confirm that the acceleration variance of the $45 \mu
\mathrm{m}$ particles is within a few percent of the zero particle
size limit.
The inset shows the ratio of the $a_0$ values for transverse
and axial components of the acceleration.}
\label{fig:acc_Rlam}
\end{figure}

Our measurement of the Kolmogorov constant $a_0$ is shown
in Fig.~\ref{fig:acc_Rlam}
for eight orders of magnitude of scaling in acceleration variance.
We find $a_0$ to be anisotropic and to depend
significantly on the Reynolds number.
The $a_0$ values for both components increase as a function
of Reynolds number up to
$R_\lambda \approx 500$, above which they are approximately constant. 
The trend in $a_0$ is
consistent with DNS results in the range $140 \le R_\lambda \le
470$\cite{vedula:1998,gotoh:1999,gotoh:2000}.
However, the constant value of $a_0$ at high Reynolds number suggests that
K41 scaling becomes valid 
at higher Reynolds numbers.
Weak deviations from the
K41 scaling such as the $a_0 \sim R_\lambda^{0.135}$ prediction of the 
multifractal model by Borgas\cite{Borgas:1993:MLN} cannot be ruled out by our
measurements.

The acceleration variance is larger for the transverse
component than for the axial component at all values of the Reynolds number.
This is shown in the inset to Fig.~\ref{fig:acc_Rlam} 
where the ratio of the Kolmogorov
constants for the axial and transverse acceleration components is plotted as
a function of Reynolds number. 
The anisotropy is large at low Reynolds number and diminishes to
a small value at $R_\lambda = 970$.
This observation tends to confirm recent experimental results which
indicate that anisotropy may persist to much higher Reynolds
numbers than previously believed\cite{sreeni:2000,warhaft:2000}.

In summary, our measurements indicate that the Heisenberg-Yaglom 
scaling of acceleration variance is observed 
for $500 \le R_\lambda \le 970$.
At lower Reynolds number, our measurements are consistent
with the anomalous scaling observed in DNS\cite{vedula:1998,gotoh:2000}.
Our measurements show that the anisotropy of the large scales
affects the acceleration 
components even at $ R_\lambda \approx 1000$. 
It is impossible to say on the basis of these measurements
if the anisotropy will persist as the Reynolds number approaches
infinity.
We found the acceleration distribution to be very intermittent, with extremely
large accelerations often arising in vortical structures 
such as the one shown in
Fig.~\ref{fig:traj}.

Our results have immediate application for the development of Lagrangian
stochastic models, some of which use $a_0$ directly as a model constant. 
These models are being developed and used to efficiently simulate mixing,
particulate transport, and combustion in practical flows with varying Reynolds
numbers\cite{pope:1994,reynolds:1999,sawford:2000b}. 
Our research also has surprising implications for everyday phenomena.
For instance, a mosquito flying on a windy day (wind speed 18~km/h 
and an altitude of 1 meter) would experience an rms acceleration of
$15\ \mathrm{m}/\mathrm{s}^2$.  But given the extremely intermittent
nature of the acceleration, our mosquito could expect to experience
accelerations of $150\ \mathrm{m}/\mathrm{s}^2$ (15 times the acceleration
of gravity) every 15 seconds.
This may explain why, under windy conditions, 
a mosquito would prefer to cling to a blade of grass
rather than take part in the roller coaster ride through 
the Earth's turbulent boundary layer\cite{bidlingmayer:1995}.

\newpage

\section{Acknowledgments}

This research is supported by the Physics Division of the
National Science Foundation.
We thank Reginald Hill, Mark Nelkin, Stephen B. Pope, Eric Siggia, and
Zellman Warhaft
for stimulating discussions and suggestions throughout the project.
We also thank Curt Ward, who assisted in the initial development
of the strip detector.  EB and ALP are grateful for support from
the Institute of Theoretical Physics at the University of California,
Santa Barbara, where parts of the manuscript were written.

% FIGURES

% \pagebreak
% \pagestyle{empty}

% \bibliography{turbulence2,turbulence,acc_paper,amc}
\bibliographystyle{unsrt}

\newpage
\rule{1pt}{0pt}
\newpage

\epsfxsize=\linewidth
\begin{center}
\mbox{\epsffile{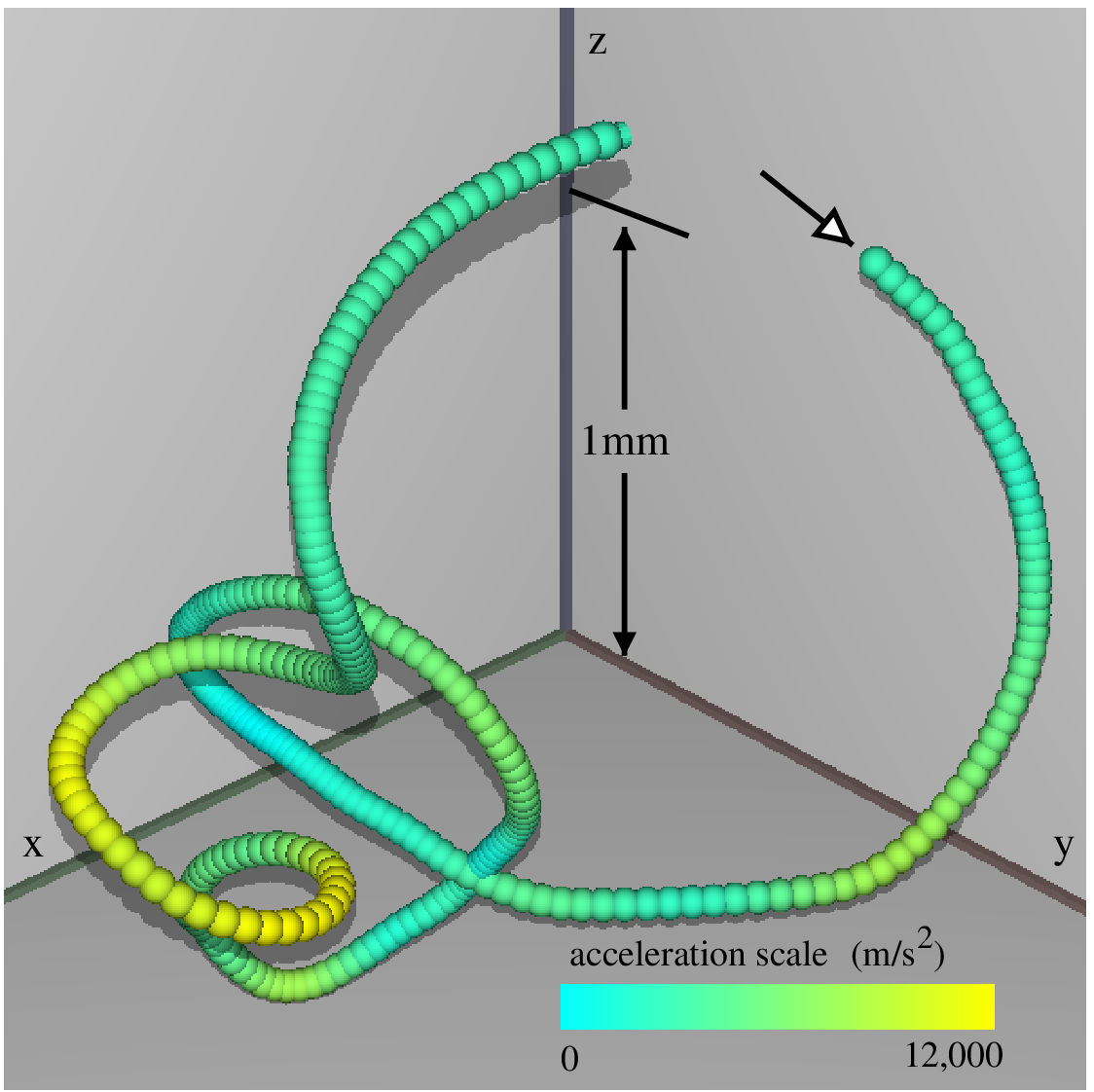}}
\end{center}

\begin{figure}[tb]
\caption{\textsc{Measured Particle Trajectory}
The 3-dimensional time-resolved trajectory of a 46 micrometer  
diameter particle in a turbulent water flow 
at Reynolds number 63,000 ($R_\lambda = 970$).
A sphere marks the measured position of the particle in each of 300 frames
taken every 0.014~ms ($\approx \tau_\eta/20$).
The shading
indicates the acceleration magnitude, with the maximum value of
12,000~m/$\mathrm{s}^2$ corresponding to 
approximately 30 standard deviations. 
The turbulence is generated between coaxial counter-rotating
disks\protect\cite{Voth:1998:LAM,laporta:1999}
in a closed flow chamber of volume 0.1~$\mathrm{m}^3$ with
rotation rates ranging from 0.15~Hz to 7.0~Hz, giving
rms velocity fluctuation $\tilde u$ in the range 
$0.018 \mathrm{m}/\mathrm{s} < \tilde u <
0.87 \mathrm{m}/\mathrm{s}$.
Measurements are made in an 8~$\mathrm{mm}^3$ volume
at the center of the apparatus where the mean velocity is zero
and the flow is nearly homogeneous
but not isotropic.
As a result of a mean stretching of the flow along the propeller axis
the rms fluctuations are 1/3 larger for the transverse velocity components
than for the axial component.
The energy dissipation was determined from measurements
of the transverse second order structure function and
the Kolmogorov relation 
$D_{NN} = \frac{4}{3} C_1 \left( \epsilon r \right)^{2/3}$
with $C_1 = 2.13$\protect\cite{sreeni:1995}.
The dissipation was found to be related to the rms velocity fluctuation
by $\epsilon = {\tilde u}^3/L$ 
with an energy injection scale $L = (71 \pm 7)\ \mathrm{mm}$. 
Using the definition of the Taylor microscale Reynolds number
$R_\lambda = \left(15 \tilde u L / \nu\right)^{1/2}$ the 
range of Reynolds numbers accessible is $140 \le
R_\lambda \le 970$, (in terms of the classical Reynolds number
$1300 \le \mathrm{Re} \le 63,000$). 
At the highest Reynolds number the system is characterized
by Kolmogorov distance and time scales of 
$\eta = 18\ \mu\mathrm{m}$ and
$\tau_\eta = 0.3\ \mathrm{ms}$, 
respectively.} 
\label{fig:traj}
\end{figure}

\end{document}